\documentclass[12pt]{article}
\usepackage{amssymb}
\usepackage{amsmath}
\usepackage{cprotect}
\usepackage{cite}
\usepackage{hyperref}
\usepackage{amsfonts}
\usepackage{bbm}
\usepackage{physics}
\usepackage{tikz-cd}
\usepackage{MnSymbol}
\input xy
\xyoption{all}
\usepackage{enumitem}

\usepackage{graphicx} 

\numberwithin{equation}{section}

\numberwithin{table}{section}
\newcommand{\dilog}{\mathrm{Li_2}}

\begin{document}

\begin{center}

{\large\bf Towards quantum elliptic cohomology from GLSMs}

\vspace*{0.2in}

Z.~Cao$^1$, E.~Sharpe$^2$, H.~Zhang$^2$

\begin{tabular}{cc}
{\begin{tabular}{l}
$^1$ Center for Cosmology \\
$\quad$ and Particle Physics\\
New York University\\
New York, NY 10003
\end{tabular}}
&
{\begin{tabular}{l}
$^2$ Department of Physics MC 0435 \\
850 West Campus Drive \\
Virginia Tech \\
Blacksburg, VA 24061 
\end{tabular}}
\end{tabular}

\vspace*{0.2in}

{\tt zc3575@nyu.edu}, {\tt ersharpe@vt.edu}, {\tt hzhang96@vt.edu}

\end{center}

This paper briefly outlines a proposal for a GLSM Coulomb-branch interpretation of the quantum elliptic cohomology rings currently being developed in mathematics by Bouaziz-Huq-Kuruvilla-Lee.  The proposal is modeled on the established GLSM computations of quantum cohomology and quantum K theory of Fano toric varieties, where ring relations arise as critical loci of twisted effective superpotentials.
In the case of quantum elliptic cohomology, the natural generalization would be to encode ring relations in Coulomb branch computations in two-dimensional theories obtained by 
KK reductions from four-dimensional theories, describing OPEs of parallel surface defects, which we review.  However, in some basic examples, this interpretation is obstructed by four-dimensional gauge anomalies.  In such cases, we provide a classical superpotential whose critical loci generate the expected ring relations, but which is not obtained by KK reduction from four dimensions, though it shares many characteristics with such KK reductions.  We describe some simple examples, and make some predictions for the case of gerbes.

\begin{flushleft}
    September 2026
\end{flushleft}

\newpage

\tableofcontents

\newpage

\section{Introduction}

The purpose of this paper is to briefly outline a proposal for how the ``quantum elliptic cohomology'' rings under development in mathematics  \cite{ypleetoappear,ypleetalks} can be computed using gauged linear sigma model (GLSM) \cite{Witten:1993yc} Coulomb branch techniques.
The idea is that, at least morally, they should arise from 
parallel surface operators wrapped on two-tori, computed explicitly via KK reduction of four-dimensional gauge theories to two dimensions.  That said, we will see that the two-dimensional theory that seems to be relevant will be distinct from, though closely related to, the KK reduction in typical examples.

The prototypes for these computations are quantum cohomology and quantum K theory computations in GLSMs.
Consider a GLSM for a Fano GIT quotient $V//G$ for $V$ a vector space.
RG flow drives the GLSM from a Higgs branch realizing the geometry, onto a Coulomb branch, which is not geometric.  As topological field theory correlation functions are invariant under RG flow, the quantum cohomology of the geometry (Higgs branch) can be computed along the Coulomb branch, where the relations are given as the critical locus of a twisted one-loop effective superpotential in a two-dimensional GLSM \cite{Morrison:1994fr}.

The quantum K theory ring relations of $V//G$ can be computed similarly.
Here, one considers a three-dimensional GLSM for $V//G$, on a three-manifold $S^1 \times \Sigma$ for $\Sigma$ a Riemann surface, and KK reduces along the $S^1$ to a two-dimensional theory with a resummed infinite tower of massive modes.
Along the Coulomb branch, again one computes the critical locus of a twisted one-loop effective superpotential,
which is now interpreted physically as OPEs of line operators wrapped on the $S^1$, and mathematically as quantum K theory ring relations \cite{Bullimore:2014awa,Jockers:2018sfl,Jockers:2019wjh,Jockers:2019lwe,Jockers:2021omw,Ueda:2019qhg,Koroteev:2017nab,Gu:2020zpg,Gu:2023tcv,Gu:2022yvj,Gu:2023fpw,Sharpe:2024ujm,Huq-Kuruvilla:2025nlf,epiga1,irit1}.

In principle, the idea of this paper is to apply the same methods to KK reductions of four-dimensional gauge theories on four-manifolds of the form $T^2 \times \Sigma$, KK reducing along $T^2$ and computing Coulomb branch superpotential critical loci.  Physically, this should compute OPEs of parallel surface operators wrapped on the $T^2$, and mathematically as quantum elliptic cohomology.  That said, the prototypical examples of which we are aware \cite{ypleetoappear,ypleetalks} correspond to four-dimensional gauge theories with gauge anomalies, and in such examples the procedure just outlined is not defined.  
Instead, in such cases, we provide a classical superpotential whose critical loci appear to generate the quantum elliptic cohomology relations.  This superpotential cannot arise from a KK reduction from four dimensions, though it shares many characteristics with such.  (In such cases, whether that superpotential arises from a sensible supersymmetric QFT is a question left for future work.)

To be clear, reductions of gauge theories from four dimensions to two dimensions have appeared in many
other places in the literature, see for example \cite{Harvey:1995tg,Bershadsky:1995vm,Kapustin:2006pk,Kutasov:2013ffl,Closset:2017bse,Hwang:2018riu,Nekrasov:2009uh,Dedushenko:2021mds,Dedushenko:2023qjq,Longhi:2019hdh,NoumanMuteeb:2025hkx,Ding:2025ynw}.  Furthermore, ordinary elliptic cohomology arising from interfaces and defects of three- and four-dimensional theories on $T^2$
has been discussed in, for example, \cite{Aganagic:2016jmx,Dedushenko:2021mds,Dedushenko:2023qjq,Bullimore:2021rnr,Ishtiaque:2023acr,frv,os}.
See in particular, for example, \cite[section 3.3]{Dedushenko:2021mds} for a clear elaboration of why one expects elliptic cohomology to arise in reductions on $T^2$.  
(Ordinary) elliptic cohomology itself has been discussed in a number of references, see for
example \cite{gkv,groj,gm,st,bouaziz-khan} or the surveys \cite{lurie1,be}.  For the purposes of this paper, we will describe elliptic cohomology as merely, power series in K theory classes.

The purpose of this paper is to propose a GLSM Coulomb branch interpretation of the quantum elliptic cohomology under development by Bouaziz-Huq-Kuruvilla-Lee \cite{ypleetoappear,ypleetalks}, for which we will see that a two-dimensional theory slightly different from a standard KK reduction is needed.

We begin in section~\ref{sect:rev} by reviewing KK reductions from four dimensional gauge theories to two dimensions.
In section~\ref{sect:qec} we describe how a closely-related two-dimensional theory, which is not a KK reduction, can reproduce the ring relations of \cite{ypleetoappear}.
In section~\ref{sect:gerbes} we outline expectations for the case that the GLSM is describing a (Deligne-Mumford) gerbe.  
Because of decomposition \cite{Hellerman:2006zs}, quantum cohomology of gerbes is equivalent to that of a disjoint union of spaces,
and in quantum K-theory computations, one gets a squared decomposition \cite{Gu:2021yek,Gu:2021beo,Sharpe:2024ujm,Huq-Kuruvilla:2025nlf}.
We outline the analogous expectation here, predicting a cubed decomposition.
Finally, in section~\ref{sect:surface}, we make some general observations about surface defects and elliptic genera.

This paper began two years ago as a summer project with a visiting undergraduate, the first author, currently a graduate student at NYU.
While this work was under development, we learned of the upcoming publication \cite{hanstoappear},
which prompted us to write up our limited current results and coordinate submissions.

\section{Review of KK reduction in GLSMs}  \label{sect:rev}

Just as quantum K theory is computed in GLSMs by OPEs of parallel Wilson lines in a three-dimensional
theory compactified on an $S^1$, here we will be interested in computing OPEs of parallel surface operators compactified to two dimensions on a $T^2$.   By close analogy with the three-dimensional case, these will be computed via Kaluza-Klein reduction of a four-dimensional theory.  In the resulting $2d$ theory, the infinite tower of Kaluza-Klein modes on $T^2$ will result in modification of the twisted superpotential  \cite{Witten:1993yc,Morrison:1994fr} via one-loop corrections to the Fayet-Iliopoulos parameter and the theta angle.

\subsection{Four-dimensional gauge anomalies}  \label{sect:anom}

We are working in a Kaluza-Klein reduction along $T^2$ of a four-dimensional $N=1$ super Yang-Mills theory.
If that four-dimensional theory is a quantum field theory,
instead of a classical theory, then gauge anomalies of the four-dimensional theory will
constrain possible matter content, hence target spaces of the two-dimensional theory.  
Let us quickly walk through the details.

For simplicity, we will specialize to abelian gauge theories.  Let $T$ be the gauge group, and $Q_i^a$ the charge of the $i$th component of the
matter representation with respect to the $a$th $U(1) \subset T$.  These are interpreted mathematically as the Chern roots of the tangent bundle $TM$ of the Higgs branch -- here, the target-space $M$ of the two-dimensional theory.
The linear and cubic anomalies in the four-dimensional theories give\footnote{
Recall that there is no quadratic anomaly, as it would couple to the trace of the spacetime curvature, which as an $SO(n)$ matrix, is traceless.
} the constraints
\begin{equation}
    \sum_i Q_i^a \: = \: 0, \: \: \:
    \sum_i Q_i^a Q_i^b Q_i^c \: = \: 0.
\end{equation}

Applying the splitting principle, and suppressing the $a$ indices, one can write
\begin{eqnarray}
    c_1(TM) & = & \sum_i Q_i,
    \\
    c_2(TM) & = & \sum_{i < j} Q_i Q_j,
    \\
    c_3(TM) & = & \sum_{i < j < k} Q_i Q_j Q_k,
\end{eqnarray}
from which we more or less immediately see that gauge anomaly cancellation in the four-dimensional theory
seems to imply $c_1(TM) = 0$.  Furthermore, using the fact that
\begin{eqnarray*}
    ch_3(TM) & = & \frac{1}{3!} \sum_i Q_i^3,
    \\
    & = &
    \frac{1}{6} c_1(TM)^3 + \frac{1}{2} c_3(TM) - \frac{1}{2} c_1(TM) c_2(TM),
\end{eqnarray*}
we see that altogether the gauge anomaly cancellation conditions imply
\begin{equation}  \label{eq:anom-constr}
    c_1(TM) \: = \: 0 \: = \: c_3(TM).
\end{equation}

\subsection{Twisted effective superpotential}

 Following\footnote{
 The literature has many other examples of reductions from four dimensions to two, as partially listed in the introduction.
 } the four-dimensional reduction given in \cite[section 2.2.3]{Nekrasov:2009uh}, 
  \cite[section 2.1]{Closset:2017bse},
in more nearly the conventions of the quantum K theory computations described in \cite[section 2.1]{Gu:2020zpg},
for an anomaly-free abelian four-dimensional theory,
we take
the regularized effective twisted superpotential to have the form
\begin{equation}  \label{twisted_superpotential}
\begin{aligned}
    \widetilde{W}  =  \frac{1}{2}&k^{ab}\qty(\ln{X_a})(\ln{X_b}) +  \frac{1}{4}\sum_i\qty(\ln{\qty(X^{\rho_i} y_i)})^2 + \sum_a(\ln{\widetilde{q}_a})(\ln{X_a})
    \\
    & - \frac{1}{12}\sum_i \qty(\ln{q}) \qty(\ln{X^{\rho_i}}) - \sum_{n=0}^\infty \sum_i \qty[\dilog \qty(q^{n+1}X^{-\rho_i} y_i^{-1})-\dilog\qty(q^{n}X^{\rho_i} y_i)],
    \end{aligned}
\end{equation}
where
\begin{equation}
    \widetilde{q}_a = \exp( 2\pi i \hat{t}_a) , \qquad 
    q = \exp(2\pi i \tau),
\end{equation}
and
\begin{equation}
    X_a = \exp(2\pi i R \sigma_a),\qquad y_i = \exp(2\pi i R \widetilde{m}_i).
\end{equation}
Here, $\hat{t}_a$ is the complex coupling for $4d$ gauge fields and $\tau$ is the modular parameter of the internal 2-torus.
As before, for simplicity we specialize to abelian gauge theories.
Twisted masses for the chiral multiplets  are encoded in $y_i$, given by the following form of the Lagrangian\cite{Gates:1983py,Gates:1984nk,Nekrasov:2009uh}.
\begin{equation}
    \int \dd^4 \theta \Tr_{\mathcal{R}}{\Phi ^\dagger \qty(\sum_i e^{\widetilde{V_i}}\otimes \mathrm{Id}_{R_i})} \Phi,
    \quad \widetilde{V_i}=\widetilde{m}_i \theta_+ \bar{\theta}_-.
\end{equation}

We have included a few deformations, in the spirit of quantum K theory computations,
to make the twisted effective
superpotential~(\ref{twisted_superpotential}) more general, so strictly speaking what we have written is not solely the result of a KK reduction. (It is better described as a KK reduction along one $S^1$, followed by adding some three-dimensional Chern-Simons terms, followed by a second KK reduction.)
In analogous three-dimensional computations, quantum K theory is obtained only if one uses one particular set of Chern-Simons levels, determined by ``$U(1)_{-1/2}$ quantization,'' see e.g.~\cite[section 2.2]{Closset:2019hyt}, \cite[section 2.2]{Gu:2020zpg}.  Similarly here, our expectation is that only one choice of deformation parameter will duplicate quantum elliptic cohomology.  We leave a systematic examination of such choices for future work.

In passing, we note that in the limit $q \rightarrow 0$ (see e.g.~\cite[section 3]{Hwang:2018riu}, \cite[section 5]{NoumanMuteeb:2025hkx}), if we suitably absorb $\ln q$ into a rescaling of $\ln \widetilde{q}$, then for abelian gauge theories this reproduces the superpotential used by two of the authors in quantum K theory
\cite[equation (2.1)]{Gu:2020zpg}, as arising in three-dimensional GLSMs.

Later in computing critical loci, it will be important to recall
\begin{equation}
    x \frac{d}{dx} {\rm Li}_2(x) \: = \: {\rm Li}_1(x) \: = \: - \ln(1-x).
\end{equation}

At least in simple cases, in which the effective two-dimensional theory flows onto a pure Coulomb branch,
the desired OPE ring relations are given as the critical locus of the effective superpotential~(\ref{twisted_superpotential}) above.  (The same is also true of quantum K theory and quantum cohomology computations in GLSMs -- at least in cases in which the (effective) two-dimensional theory flows onto a pure Coulomb branch, the OPE ring relations are also given as the critical locus of a twisted one-loop effective superpotential.)  
In the case of quantum cohomology in two dimensions, the critical locus is interpreted as OPEs of local operators;
in the case of quantum K theory arising from KK reduction of three-dimensional theories, the critical locus is interpreted as OPEs of line operators wrapped on the $S^1$ on which the KK reduction has taken place.  Here, we similarly interpret the critical locus as OPE relations for products of surface operators, wrapped on the $T^2$ on which one has performed the KK reduction.

Although it will not be relevant here, for completeness we mention that
for non-Abelian theories with gauge group $U(k)$, non-vanishing vevs in the Coulomb branch break the gauge group to $U(1)^k$. The charges of the $i$-th chiral multiplet under these $U(1)$'s are labeled $\rho^a_{ib} = (\rho_i)^a_b$, and are the weight vectors of the correspoinding representation of the gauge group. By abuse of notation, we use the following convention
\begin{equation}
    \sum_i f\qty(X^{\rho_i})=\sum_i \sum_a f\qty( \prod_b 
 X_b^{\rho^b_{ia}}).
\end{equation}
The same conventions are used in e.g.~\cite{Gu:2020zpg,Gu:2023tcv,Gu:2022yvj,Gu:2023fpw}, and further details can be found there.

Now, it is important to emphasize that in order for the four-dimensional theory to make sense,
the gauge anomaly acts as a strong restriction.  However, the effective two-dimensional theories we manipulate are defined in greater generality.  As a result, we will often manipulate effective two-dimensional theories, which morally are obtained from a four-dimensional theory, though at times the four-dimensional theory will not exist.

To make the distinction more clear, suppose, for the sake of argument, that the four-dimensional gauge anomaly cancellation condition is not obeyed, and one tries to perform a KK reduction on some torus of
characteristic scale $R$.
\begin{itemize}
    \item If the two-dimensional theory is considered at an energy scale much less than $1/R$,
    then none of the KK modes are visible, and the theory is truly two-dimensional.  This is a dimensional reduction, rather than a Kaluza-Klein reduction, and it defines a related but inequivalent theory.  In this case, the two-dimensional theory, obtained by dimensional reduction, is consistent, but is also fundamentally distinct from the four-dimensional theory of which it is a truncation.  In particular, this two-dimensional theory is well-defined regardless of any gauge anomalies in the four-dimensional theory.

    \item In this paper, we are interested in computations at scales much larger than $1/R$
    (but less than the scale at which the gauge symmetry is broken to a maximal torus),
    because we want to consider the role of the tower of KK modes in the two-dimensional theory.

    Here, we cannot ignore the four-dimensional gauge anomaly.  By keeping the entire\footnote{
    In fairness, the contributions from the infinite tower of KK modes are regularized;
    nevertheless, for example \cite{Closset:2017bse} is able to compute the action of modular transformations on the internal $T^2$ on the low-energy superpotential, strongly suggesting that the effects of the regularization on the four-dimensional interpretation are minimal.
    } KK tower of modes, our two-dimensional theory is effectively four-dimensional, and a gauge anomaly in four dimensions should also render the two-dimensional gauge theory inconsistent.

    We can see that inconsistency explicitly as follows.
The references \cite{Corvilain:2017luj}, \cite[section 2.1]{Cheng:2021zjh} consider the closely related problem of KK reducing a four-dimensional theory with a gauge anomaly to a three-dimensional one.
They argue that after KK reduction from four-dimensions, the three-dimensional theory picks up a Chern-Simons term which explicitly breaks the gauge symmetry.
Similarly, after further KK reduction to two dimensions, one again expects non-gauge-invariant couplings that explicitly break the gauge symmetry.

See also \cite[section 4.7]{Longhi:2019hdh} for another perspective on anomalies and partition functions.
\end{itemize}

We emphasize these points because in section~\ref{sect:qec}, we will encounter a two-dimensional theory with what appears to be a tower of KK modes from a four-dimensional theory, but for which the four-dimensional theory has gauge anomalies, rendering it inconsistent. That two-dimensional theory, despite appearances, cannot be interpreted in terms of a KK reduction from four dimensions, for the reasons outlined above.

\subsection{Twisted masses}

The fact that we are constrained to Calabi-Yau's appears to pose a problem for our attempted computation, as GLSM Coulomb branch methods do not ordinarily apply to Calabi-Yau's.
The reason for this is elementary -- although a twisted one-loop effective superpotential still exists in Calabi-Yau cases, all dependence on $\sigma$ fields cancels out, and instead one gets an equation specifying the singular locus in the space of FI parameters.  Similarly, in reductions from three and four dimensions, in Calabi-Yau cases, one cannot get nontrivial cohomological (or K-theoretic) relations.

Now, this story is modified in the presence of twisted masses / equivariant parameters, and in the presence of such, it becomes possible to get cohomological constraints.

First, recall twisted masses in GLSMs, as discussed in e.g.~\cite[section 3]{Hori:2000kt}.  As described there, twisted masses can be interpreted as the lowest component of the field-strength superfield for a flavor symmetry.  Recall that the field strength superfield $\Sigma \propto \overline{D}_+ D_- V = \sigma + \cdots$, so twisted masses represent a modification of $\sigma$.
For example, they should change the interaction terms
\begin{equation}
    \sum_{i,a} Q_{i,a} \left( \overline{\sigma}_a \overline{\psi}_{+ i} \psi_{- i} +
    \sigma_a \overline{\psi}_{- i} \psi_{+ i} \right)
\end{equation}
to
\begin{equation}
    \sum_i \left[ \left( \sum_a Q_{i,a} \overline{\sigma}_a + \overline{m}_i \right) \overline{\psi}_{+ i} \psi_{- i} + \left( \sum_a Q_{i,a} \sigma_a + m_i \right) \overline{\psi}_{- i} \psi_{+ i} \right]
\end{equation}
and the interaction terms
\begin{equation}
    \sum_{i} \left| \sum_a Q_{i,a} \sigma_a   \right|^2 | \phi_i |^2
\end{equation}
become
\begin{equation}
    \sum_i \left| \sum_a Q_{i,a} \sigma_a + m_i \right|^2 | \phi_i |^2
\end{equation}

Note that these are automatically gauge-invariant -- gauge invariance does not impose any constraints on the twisted masses.

In the Calabi-Yau case, the reason one can get a cohomology relation and not just a constraint showing where the Coulomb branch attaches, is that for generic twisted masses, there's no solution for a point of intersection of the Coulomb branch with the Higgs branch.  

For example, one can recover equivariant quantum cohomology relations for a Calabi-Yau in this way.
Consider for example the total space of ${\cal O}(-2) \rightarrow {\mathbb P}^1$.  For generic twisted masses, the Coulomb branch relation is
\begin{equation}
    \left( \sigma - m_1 \right) \left( \sigma - m_2 \right) \: = \: q \left( -2 \sigma - m_p \right)^2 ,
\end{equation}
which is precisely \cite{sheldonpriv} the equivariant quantum cohomology ring relation, for generic twisted masses.
In the limit that all the twisted masses vanish, the $\sigma$'s cancel out, and this becomes a constraint on $q$, as expected.

\subsection{Example: $T^* {\mathbb P}^n$}

As a concrete example, we consider the case that $M = T^* {\mathbb P}^n$.  As a toric variety, this is described by
a $U(1)$ gauge theory with $2n+2$ chiral superfields, half of charge $+1$ and half of charge $-1$.
It is straightforward to check
that $c_1(M) = 0 = c_3(M)$, so this is an anomaly-free case.

In this case, the twisted one-loop effective superpotential~(\ref{twisted_superpotential}) specializes to
\begin{eqnarray*}
    \widetilde{W} & = &
    \frac{1}{2} k (\ln X)^2 \: + \:
    \frac{1}{4} \sum_{i=0}^n \left( \ln(X y_i) \right)^2 \: + \:
    \frac{1}{4} \sum_{i=0}^n \left( \ln (X^{-1} y_{i+n+1}) \right)^2 \: + \:
     (\ln \widetilde{q}) ( \ln X )
    \\
    & & 
    \: - \: \sum_{i=0}^n \sum_{m=0}^{\infty} \left(
    {\rm Li}_2( q^{m+1} X^{-1} y_i^{-1}) - {\rm Li}_2( q^m X y_i)
    + {\rm Li}_2(q^{m+1} X y_{i+n+1}^{-1}) - {\rm Li}_2( q^m X^{-1} y_{i+n+1})
    \right).
\end{eqnarray*}

It is straightforward to check that the critical locus is given by
\begin{eqnarray*}
    X^{-k} X^{-(n+1)} \left( \prod_{i=0}^n \left( \frac{y_i}{y_{i+n+1}} \right)^{-1/2} \right)
    \prod_{i=0}^n \prod_{m=0}^{\infty} \frac{
    (1 - q^{m+1} X^{-1} y_i^{-1} ) ( 1 - q^m X y_i)
    }{
    (1 - q^{m+1} X y_{i+n+1}^{-1} ) ( 1 - q^m X^{-1} y_{i+n+1} )
    }
    & = & \widetilde{q}.
\end{eqnarray*}

At least morally, these should be interpreted as ring relations for OPEs of colliding surface operators,
in the same way that KK reductions of three-dimensional GLSMs are interpreted as OPEs of colliding Wilson lines.

The reader should also note that in the case when all $y_i = 1$ (meaning, all twisted masses vanish),
if we take $k = 0$, then the
$X$'s all cancel out, leaving a constraint on $\widetilde{q}$, just as in quantum cohomology Coulomb branch computations for Calabi-Yau's.

\section{Quantum elliptic cohomology}  \label{sect:qec}

In this section, we will outline how to interpret these computations mathematically in terms of the quantum elliptic cohomology currently under development,
following e.g.~\cite{gkv,bouaziz-khan,ypleetoappear,ypleetalks}.

Briefly, for our purposes, the elliptic cohomology of a space $M$ means convergent power series in
\begin{equation}
K(M)[[q]]
\end{equation}
where $q$ is related to the modular parameter $\tau$ of a fixed torus.
Define 
\begin{equation}  \label{eq:theta-defn}
    \theta(z) \: = \: \prod_{n \geq 0} \left( (1 - q^n z) (1- q^{n+1} z^{-1}) (1 - q^{n+1}) \right),
\end{equation}
which has the useful property that $\theta(z^{-1}) = - z^{-1} \theta(z)$.
Then, the classical, non-equivariant, elliptic cohomology ring of ${\mathbb P}^n$ is proposed to be given by \cite{ypleetoappear,ypleetalks}
\begin{equation}
    {\mathbb C}[ \theta( {\cal O}(1)) ] / \left( \theta({\cal O}(1))^{n+1} \right),
\end{equation}
where $\theta( {\cal O}(1))$ is understood by formally expanding the $\theta$ of equation~(\ref{eq:theta-defn}) in a power series valued in K theory.
The quantum elliptic cohomology ring of ${\mathbb P}^n$ is
proposed in \cite{ypleetoappear} to be
\begin{equation}  \label{eq:qec-pn}
    {\mathbb C}[ \theta({\cal O}(1)) ] / \left( \theta({\cal O}(1))^{n+1} - Q \right),
\end{equation}
where $Q$ is a Novikov variable.

To compare, recall the quantum cohomology ring of ${\mathbb P}^n$ has the form
\begin{equation}
    {\mathbb C}[x] / \left( x^{n+1} - Q \right),
\end{equation}
for $x$ a generator of degree-two cohomology, and the quantum K theory ring of ${\mathbb P}^n$ has the form
\begin{equation}
    {\mathbb C}[X] / \left( (1-X)^{n+1} - Q \right),
\end{equation}
for $X \sim {\cal O}(-1)$, so the structure of the quantum elliptic cohomology ring is very similar to both of these rings.

Next, we will observe that in the case of ${\mathbb P}^n$, the superpotential~(\ref{twisted_superpotential}) generates the quantum elliptic cohomology ring relations from its critical locus.  
To be clear, the case ${\mathbb P}^n$ is described by an anomalous four-dimensional GLSM,
whereas the superpotential~(\ref{twisted_superpotential}) was obtained by KK reduction in non-anomalous cases.
As a result, we cannot interpret the resulting superpotential as arising from the KK reduction of a four-dimensional theory, and as it appears to describe a $T^2$'s worth of resummed KK modes, it is not clear whether the superpotential in this case can arise from a supersymmetric QFT at all.  Nevertheless, the fact that it generates mathematical ring relations is of interest, and the question of the existence of a corresponding QFT in such anomalous cases we leave for the future.

For the case of ${\mathbb P}^{n}$, the expression for the superpotential~(\ref{twisted_superpotential})
reduces to
\begin{eqnarray}
    \widetilde{W} & = &
    \frac{k}{2} (\ln X)^2 \: + \:
    \frac{n+1}{4}( \ln X)^2 \: + \:
    (\ln \widetilde{q}) (\ln X) \: - \:
    \frac{n+1}{12} (\ln q) (\ln X)
    \nonumber \\
    & & \: - \:
     \sum_{m=0}^{\infty} (n+1) \left( {\rm Li}_2\left( q^{m+1} X^{-1} \right) -
    {\rm Li}_2\left( q^m X \right) \right).
\end{eqnarray}

The critical locus can easily be shown to be
\begin{equation}
    X^{-k - (n+1)/2} \left[ q^{1/12} 
    \prod_{m=0}^{\infty} \left(1 - q^m X\right) \left(1 - q^{m+1} X^{-1} \right) \right]^{n+1}
    \: = \: \widetilde{q}.
\end{equation}
We will take the quantity $k$, which arose as a deformation, to have the value 
$k = - (n+1)/2$, which will have the cleanest mathematical interpretation.
(This is the same as the $U(1)_{-1/2}$ quantization value for $k$, as appears in relating three-dimensional KK reductions
to ordinary quantum K theory, see e.g.~\cite[section 2.2]{Closset:2019hyt}, \cite[section 2.2]{Gu:2020zpg}.  Whether $U(1)_{-1/2}$ quantization is as relevant for quantum elliptic cohomology as it is for quantum K theory, is left for future work.)

Using the definition of the 
Dedekind eta function
\begin{eqnarray}
    \eta(\tau) & = & q^{1/24} \prod_{m=1}^{\infty} \left( 1 - q^m \right),
\end{eqnarray}
for $q = \exp(2 \pi i \tau)$,
we see that for $k = - (n+1)/2$, the critical locus equation can be written
\begin{equation}  \label{eq:pn}
    \left( q^{+1/8} \frac{ \theta(X) }{ \eta(\tau) } \right)^{n+1} \: = \: \widetilde{q}.
\end{equation}
This matches the mathematical proposal~(\ref{eq:qec-pn}), namely $\theta(X)^{n+1} = Q$, if we identify $Q = \widetilde{q} q^{-(n+1)/8} \eta(\tau)^{n+1}$,
and $X \sim {\cal O}(-1)$, much the same way as quantum K theory arises from Coulomb branch computations in GLSMs \cite{Gu:2020zpg,Gu:2023tcv,Gu:2022yvj,Gu:2023fpw,Sharpe:2024ujm,Huq-Kuruvilla:2025nlf,epiga1}.

In passing, we note that the left-hand-side of equation~(\ref{eq:pn}) is not modular-invariant.
The expression $q^{+1/8} \theta(X)/\eta(\tau)$ coincides, up to a factor of $x^{-1/2}$, with the
$\theta(x)$ in \cite[equation (B.6)]{Closset:2017bse}, for which modular transformations are nontrivial and given explicitly
in \cite[equation (B.9)]{Closset:2017bse}.  We suspect this is tied to four-dimensional anomalies, but leave
that for future work.

As a consistency check, let us take a limit to reproduce the three-dimensional theory, where analogous Coulomb branch computations give quantum K theory.
Following \cite[section 3]{Hwang:2018riu} (see also \cite[section 5]{NoumanMuteeb:2025hkx}), we take $q \rightarrow 0$.  In that limit,
$\theta(X) \rightarrow 1-X$, so that if we suitably rescale $\widetilde{q}$, the quentum elliptic cohomology
relations reduce to $(1-X)^{n+1} = Q'$ for some constant $Q'$, matching the quantum K theory relations in this case (see e.g.~\cite{Gu:2020zpg}), as expected.

That said, the fact that $Q$ is a power series in $q$, rather than a constant, may reflect the fact that this is formally a deformation of a KK reduction of an anomalous four-dimensional theory, and so might reflect deeper issues.  We remind the reader that in this section, we have taken the superpotential to be used classically, because the KK mode tower it nearly encodes would result in an anomalous four-dimensional theory.  We leave further such analysis for the future.

In passing, we note that our critical locus expression is a function of a parameter $k$, in principle arising from a deformation of the three-dimensional theory obtained halfway through the KK reduction.  In ordinary quantum K theory computations, that parameter is a Chern-Simons level, which is related to twistings of the quantum K theory,
see e.g.~\cite{Huq-Kuruvilla:2025nlf} and references therein.  It is tempting to speculate that there may exist analogous twistings of quantum elliptic cohomology.

\section{Gerbes and cubed decompositions}  \label{sect:gerbes}

Next, we turn to the special case that the target is a (Deligne-Mumford) gerbe.

If $M$ is a Deligne-Mumford gerbe, then a sigma model on $M$ is equivalent to a sigma model on a disjoint union of copies of underlying spaces, as described in detail in \cite{Hellerman:2006zs} as an example of the principle of decomposition, introduced there.  This implies that the Gromov-Witten invariants and quantum cohomology rings of such a gerbe are equivalent to those of a disjoint union of spaces, which has been checked rigorously in e.g.~\cite{ajt1,ajt2,ajt3,t1,gt1,xt1}.

There is an analogous story for quantum K theory.  Quantum K theory is realized physically via OPE rings of parallel Wilson lines on $S^1$'s, and is computed by three-dimensional sigma models on $S^1 \times \Sigma$ ($\Sigma$ a Riemann
surface), which are Kaluza-Klein reduced to $\Sigma$.  If the three-dimensional theory has a one-form symmetry
(as will be the case if $M$ is a gerbe), then the KK reduction will have both a one-form symmetry as well as a zero-form symmetry.  In the IR limit where the quantum K theory computations take place, this will manifest itself as, naively, two factors of decomposition -- one due to the presence of a one-form symmetry in two dimensions, the other due to the IR limit of a spontaneously broken symmetry.  Hence, physics predicts a `squared' decomposition.  This has been explored in detail in physics in \cite{Gu:2021yek,Gu:2021beo,Sharpe:2024ujm}, and a rigorous computation in the special case of classifying gerbes is described in \cite{Huq-Kuruvilla:2025nlf}.

Here we expect a similar story.  Begin with a four-dimensional theory with a one-form symmetry.
After a KK reduction to two dimensions, one expects the low-energy theory will have a one-form symmetry plus a pair of zero-form symmetries.  For the same reasons as above, one expects the theory to exhibit both decomposition as well as a pair of spontaneously-broken symmetries, hence deep in the IR, one expects the theory to exhibit a cubed decomposition,
so long as the four-dimensional theory is well-defined.

That said, our realization of quantum elliptic cohomology for ${\mathbb P}^n$ involved a superpotential that was not derived from a KK reduction from a four-dimensional theory, though it is closely related.  We will briefly illustrate below how a cubed decomposition arises, formally, in the example of a projective space, but as the theory we are using to describe quantum elliptic cohomology will not be a KK reduction from four dimensions if the anomaly conditions are not satisfied, the analysis above need not apply in general.

As a simple prototypical example, let us consider the ${\mathbb Z}_{\ell}$ gerbe described by the weighted projective space
${\mathbb P}^n_{[\ell,\ell, \cdots, \ell]}$, as described in \cite{Pantev:2005zs}.  Let us quickly review quantum cohomology and quantum K theory ring relations from the Coulomb branch in this case.
The quantum cohomology ring derived from the Coulomb branch, as described in \cite{Pantev:2005zs},
has the relation
\begin{equation}
    \sigma^{\ell(n+1)} \: = \: q,
\end{equation}
implying $\ell$ relations of the form $\sigma^{n+1} \propto q^{1/\ell}$, the quantum cohomology ring relation for ${\mathbb P}^n$.  Similarly, the quantum K theory ring relation
in this case is \cite{Sharpe:2024ujm}
\begin{equation}
    (1-X^{\ell})^{\ell (n+1)} \: = \: q.
\end{equation}
Taking $\ell$th roots gives
\begin{equation}
    (1-X^{\ell})^{n+1} \: \propto \: q^{1/\ell}.
\end{equation}
Using the fact that $X = \exp(2 \pi i R \sigma)$, and so is periodic, we find, for each of the roots above,
an additional $\ell$ copies of the critical locus defining the quantum K theory ring relations in ${\mathbb P}^n$.

In the present case of quantum elliptic cohomology of ${\mathbb P}^n_{[\ell,\ell,\cdots, \ell]}$,
the expression for the superpotential~(\ref{twisted_superpotential})
reduces to
\begin{eqnarray}
    \widetilde{W} & = &
    \frac{k}{2} (\ln X)^2 \: + \:
    \frac{n+1}{4}( \ln X^{\ell})^2 \: + \:
     (\ln \widetilde{q}) (\ln X) \: - \:
    \frac{n+1}{12} (\ln q) (\ln X^{\ell})
    \nonumber \\
    & & \: - \:
     \sum_{m=0}^{\infty} (n+1) \left( {\rm Li}_2\left( q^{m+1} X^{-\ell} \right) -
    {\rm Li}_2\left( q^m X^{\ell} \right) \right).
\end{eqnarray}
The critical locus can easily be shown to be
\begin{equation}
    X^{-k - \ell^2(n+1)/2} \left[ q^{1/12} 
    \prod_{m=0}^{\infty} \left(1 - q^m X^{\ell}\right) \left(1 - q^{m+1} X^{-\ell} \right) \right]^{\ell(n+1)}
    \: = \: \widetilde{q}.
\end{equation}
Taking $k = - \ell^2(n+1)/2$, in the spirit of $U(1)_{-1/2}$ quantization as discussed earlier, 
this can be written in the form
\begin{equation}
    \theta(X^{\ell})^{\ell(n+1)} = Q
\end{equation}
where $Q$ is a power series in $q$, that is proportional to $\widetilde{q}$.

The overall decomposition mentioned earlier can be seen in the fact that this expression involves an
$\ell$th tensor power of $\theta^{n+1}$; taking the $\ell$ roots gives $\ell$ copies of the relation for ${\mathbb P}^n$, modulo the dependence on $X^{\ell}$ instead of $X$.  The remaining pair of zero-form symmetries can be understood formally from the quasi-double-periodicity of $\theta$, giving rise to $\ell^2$ as many roots as one would get from the relation $\theta(X) = Q$.

As quantum elliptic cohomology is still under development at this point, we leave further investigations for future work.

\section{Surface defects and elliptic genera}  \label{sect:surface}

Most of this paper has been focused on using physics computations to understand the proposal \cite{ypleetoappear,ypleetalks} for quantum elliptic cohomology.
A different direction is to consider OPE's of surface operators in four-dimensional quantum field theories, and relations to elliptc genera.  There are already several other works in this direction, see for example
\cite{Aganagic:2016jmx,Dedushenko:2021mds,Dedushenko:2023qjq,Bullimore:2021rnr,Ishtiaque:2023acr,frv,os}, which discuss e.g.~elliptic stable envelopes arising from boundaries of three-dimensional $N=4$ supersymmetric gauge theories,
and related interface/defect considerations.  Here instead we follow the same spirit as in quantum K theory discussions, creating an effective theory of parallel surface defects by KK reducing from a four-dimensional theory to an
effective two-dimensional theory, and not interfaces or defects.
We will just make a few observations.  In this section,
we will only consider four-dimensional theories with no gauge anomalies,
meaning we restrict to theories obeying~(\ref{eq:anom-constr}).

To match other conventions, we will also work with a slightly different superpotential convention.
Specifically,
we will use a regularized twisted superpotential in the form
\begin{equation}
\begin{aligned}
    \widetilde{W}=\frac{1}{2}&k^{ab}\qty(\ln{X_a})(\ln{X_b}) + \frac{1}{4}\sum_i\qty(\ln{\qty(X^{\rho_i} y_i)})^2 - \sum_a(\ln{\widetilde{q}_a})(\ln{X_a})
    \\
    &+\frac{1}{12}\sum_i \qty(\ln{q}) \qty(\ln{X^{\rho_i}}) + \sum_{n=1}^\infty \sum_i \qty[\dilog \qty(q^{n-1}X^{-\rho_i} y_i^{-1})-\dilog\qty(q^{n}X^{\rho_i} y^i)],
\end{aligned}
\end{equation}

In terms of this superpotential, the critical loci will prototypically be described by factors of the form
\begin{equation}
    q^{-1/12} \frac{ \theta_1(z,\tau) }{\eta(\tau)} 
    \: = \:
    -i \left(z^{-1/2} - z^{+1/2}\right)\, \prod_{m=1}^{\infty} \left( 1 - z q^m \right) \left( 1 - z^{-1} q^m \right),
\end{equation}
for 
\begin{equation}
     \theta_1(\xi,q) \: = \: -i q^{1/8} \left( z^{1/2} - z^{-1/2} \right) \prod_{m=1}^{\infty}
    \left( 1 - q^m \right) \left( 1 - q^m z  \right) \left( 1 - q^m z^{-1} \right),
    \quad z = \exp(2 \xi),
\end{equation}
instead of factors involving $q^{-1/12} \theta/\eta$, as arise in the conventions of the rest of this paper.
This is of essentially the same form as in our previous discussion (but the slightly different functional dependence will be helpful for the interpretation in this section).

In the language of surface operators, we can interpret these ratios in terms of
the contribution to the $T^2$ partition function from left-moving fermions.  Following \cite[section 2.2.1]{Ando:2009av}, in the large-radius (Free field) limit, a single complex left-moving R-sector fermion graded by $z$ contributes
    \begin{equation} \label{eq:mod-contrib}
        \left(z^{-1/2} - z^{+1/2}\right)\, \prod_{m=1}^{\infty} \left( 1 - z q^m \right) \left( 1 - z^{-1} q^m \right)
    \end{equation}
    to the partition function, where the $z^{-1/2} - z^{1/2}$ factor is due to its zero modes, and the rest is from oscillator modes.  For $z = \exp(2 \pi i \nu)$, this contribution is the same as
    \begin{equation}
        +i q^{-1/12} \frac{ \theta_1(\nu, \tau) }{ \eta(q) }.
    \end{equation}

    A very similar story exists in expressions for elliptic genera in terms of JK residues.
    Specifically, the ratio $\theta_1/\eta$ is also the contribution to a JK residue integral for
    an elliptic genus due to a single (0,2) Fermi superfield, see e.g.~\cite[equ'n (2.16)]{Benini:2013nda}.

For completeness, we can rewrite~(\ref{eq:mod-contrib}) in terms of exterior products of a line bundle $L$, as
    \begin{eqnarray}
+i q^{-1/12} \frac{ \theta_1(\nu, \tau) }{ \eta(q) }
    & = &
        {\rm ch}\left( \left( (\det L)^{-1/2} - (\det L)^{+1/2} \right) \otimes \bigotimes_{m=1,2,3,\cdots} \Lambda_{-q^m}\left( L^{\mathbb C} \right) \right),
        \\
        & = & -
        {\rm ch}\left( \left( \det L \right)^{+1/2} \otimes\Lambda_{-1}(L^*) \bigotimes_{m=1,2,3,\cdots} \Lambda_{-q^m}\left( L^{\mathbb C} \right) \right),  \label{eq:omega-rep}
    \end{eqnarray}
    for $2 \pi i \nu = c_1(L)$,
    where
    \begin{equation}
        \Lambda_q(z {\cal E}) \: = \: 
        1 \oplus z q {\cal E} \oplus z^2 q^2 {\rm Alt}^2 {\cal E} \oplus \cdots
        \: = \: \Lambda_{zq}({\cal E}),
    \end{equation}
    in the notation of \cite{Ando:2009av}.

Formally, 
given $\omega$ a product of modular forms such as $\theta_1(\nu,\tau)/\eta(q)$, one could define
\begin{equation}  \label{eq:classical-trace}
    \langle \omega \rangle \: = \: \int_M \hat{A}(TM) \wedge {\rm ch}\left( \bigotimes_{n=1,2,3,\cdots} S_{q^n} \left(TM)^{\mathbb C} \right) \right) \, \wedge \omega,
\end{equation}
where $\omega$ is more usefully represented in the form~(\ref{eq:omega-rep}).
The factors of $\hat{A}(TX)$ and ch$(S_{q^n}( (TX)^{\mathbb C} )$ are merely part of the measure, in this proposal.

To interpret this, we use the fact that for line bundles $L_1$, $L_2$, 
\begin{equation}
    \Lambda_q( L_1 \oplus L_2) \: = \: \Lambda_q(L_1) \otimes \Lambda_q(L_2),
\end{equation}
(see e.g.~\cite[appendix A]{Ando:2009av}),
so we can apply the splitting principle to a bundle ${\cal E}$ and expand $\Lambda_q({\cal E})$ in terms of line bundles $L_i$ associated to Chern roots $x_i$ as
\begin{equation}
\Lambda_q( {\cal E}) \: = \: \bigotimes_i \Lambda_q(L_i).
\end{equation}
As a result, if we let $\omega({\cal E})$ denote the product of a set of
factors~(\ref{eq:omega-rep}) associated to Chern roots of a rank $r$ bundle
${\cal E}$, then
\begin{eqnarray}
    \omega({\cal E}) & = & \bigotimes_i \omega(L_i),
    \\
    & = &
    (-)^{r} {\rm ch}\left( ({\det \cal E})^{+1/2} \Lambda_{-1}( {\cal E}^*)
    \bigotimes_{m=1,2,3,\cdots} \Lambda_{-q^m}\left( {\cal E}^{\mathbb C} \right)
    \right),
\end{eqnarray}
using the fact that ch$(A \otimes B) = {\rm ch}(A) \wedge {\rm ch}(B)$ for any two
vector bundles $A$, $B$,
and then the trace~(\ref{eq:classical-trace}) becomes
\begin{equation}
    \langle \omega({\cal E}) \rangle 
    \: = \:  (-)^{r} 
    \int_M \hat{A}(TM) \wedge {\rm ch}\left( \bigotimes_{n=1,2,3,\cdots} S_{q^n} \left(TM)^{\mathbb C} \right)
    \otimes
    ({\det \cal E})^{+1/2} \Lambda_{-1}( {\cal E}^*)
    \bigotimes_{m=1,2,3,\cdots} \Lambda_{-q^m}\left( {\cal E}^{\mathbb C} \right)
    \right).
\end{equation}

To interpret that trace above,
we use the fact
that for a (0,2) NLSM with left-moving R-sector fermions describing a bundle ${\cal E}$ of rank $r$, over a space $M$ of dimension $n$,
the elliptic genus is \cite[equ'n (31)]{Witten:1986bf},
\cite[equ'n (8)]{Ando:2009av}
\begin{eqnarray}
\lefteqn{
    {\rm Tr}_{R,R} (-)^{F_R} (-)^{F_L} q^{L_0} \overline{q}^{\overline{L}_0}
    } 
    \nonumber \\
    & = &
    q^{+(1/12)(r-n)} \int_M \hat{A}(TM) \wedge {\rm ch}\Biggl(  (-)^{r/2} ({\det \cal E})^{1/2} \Lambda_{-1} ({\cal E}^*) 
    \bigotimes_{n=1,2,3,\cdots} S_{q^n}\left( (TM)^{\mathbb C} \right) 
    \bigotimes_{n=1,2,3,\cdots} \Lambda_{-q^n}\left(  {\cal E}^{\mathbb C} \right)
    \Biggr), \nonumber
    \\
    & \propto & \langle \omega({\cal E}) \rangle. \nonumber
\end{eqnarray}

As a result, if the product of the observables $\omega(L_i)$ matches $\omega({\cal E})$,
then the proposed trace~(\ref{eq:classical-trace}) is proportional to the elliptic genus of the (0,2)-supersymmetric nonlinear sigma model above, describing the
bundle ${\cal E}$ on the space $M$.

We should observe that the corresponding two-dimensional physical theory is
only well-defined if the Green-Schwarz condition
\begin{equation}
    {\rm ch}_2({\cal E}) \: = \: {\rm ch}_2(TM)
\end{equation}
is obeyed.  (Mathematically, this is related to conditions for the elliptic genus to have good modular properties.)
Furthermore, this index also famously detects supersymmetry breaking:  if $\langle \omega({\cal E}) \rangle = 0$, then there is no obstruction to
dynamical supersymmetry breaking in the physical theory.

In the special case that ${\cal E} = TM$, the expression for $\langle \omega({\cal E}) \rangle$ simplifies, using the fact that for any vector bundle ${\cal F}$,
\begin{equation}
    ( S_q {\cal F} )^{-1} \: = \: \Lambda_{-q} {\cal F}
\end{equation}
(see for example \cite[appendix A]{Ando:2009av}).
As a result, in this case,
\begin{eqnarray}
    {\rm Tr}_{R,R} (-)^{F_R} (-)^{F_L} q^{L_0} \overline{q}^{\overline{L}_0}
    & = &
     \int_M \hat{A}(TM) \wedge {\rm ch}\left(  (-)^{r/2} ({\det \cal E})^{1/2} \Lambda_{-1} ({\cal E}^*) 
    \right).
\end{eqnarray}

Consider the case that $M = {\mathbb P}^n$ (but ${\cal E}$ is not necessarily $TM$).  In this case, for $\omega(L) \propto \theta_1/\eta$, $\langle \omega^{n+1} \rangle = 0$, formally in line with earlier observations.  We can see this as follows.
\begin{itemize}
\item Each $\omega(L) \propto \theta_1(\nu,q)$ for $\nu = c_1(L)$.  However, $\theta_1(\nu,q) \propto \sin \nu$ also, and if we expand that in a Taylor series, we get a leading
factor of $\nu$.  Hence, on ${\mathbb P}^{n}$, $\omega(L)^{n+1} = 0$, just from basic differential geometry, as it is
a $(2n+2)$-form on a $2n$-dimensional space.

    \item Each factor of $\theta_1/\eta$ corresponds to an ordinary line bundle $L \rightarrow {\mathbb P}^n$.  Then, $k$ such factors
    describes a bundle ${\cal E} = \oplus^k L$.  Note that
    ch$_2({\cal E}) = (k/2!) c_1(L)^2$.  If $L = {\cal O}(1)$, then this
    satisfies Green-Schwarz on $M = {\mathbb P}^{n}$ when $k=n+1$.  (Indeed, except for the ranks, which differ, the other Chern characters of $T {\mathbb P}^{n}$ match those of ${\cal E}$.  In fact, this ${\cal E}$ is a $C^{\infty}$ deformation of an extension of $T {\mathbb P}^{n}$ by ${\cal O}$, described by a (0,2) model in which the $\sigma$ field has been deleted. 
    
    Now, the elliptic genus of that (0,2) theory vanishes \cite{ilarionpriv}, which is certainly consistent with the statement that classically, $\omega({\cal E}) = \omega(L)^{n+1} = 0$.

\end{itemize}

It is tempting to speculate on extensions of such operations to include quantum corrections.  It is natural to conjecture that this would involve replacing $M$ with the space of maps $T^2 \rightarrow M$.  It is then natural to speculate
\cite{dbe-priv} that the gauge anomaly constraint on $c_3$ might be related to a String constraint on moduli spaces via transgression.  We leave such matters for future work.

\section{Conclusions}

In this paper, we have briefly outlined how upcoming mathematical results defining quantum sheaf cohomology \cite{ypleetoappear} can be understood via GLSMs, with the ring relations emerging as the critical locus of a superpotential, in a form similar to GLSM computations of quantum cohomology and quantum K theory of Fano GIT quotients of vector spaces.  Morally, these GLSM computations ought to be understandable as OPEs of parallel surface operators, computed via a KK reduction of a four-dimensional theory; however, as we have remarked in detail, the four-dimensional theories do not exist in all cases of interest to \cite{ypleetoappear}.

To better understand these results, as a first step, further examples of quantum elliptic cohomology in mathematics will be extremely useful.  We leave a more detailed understanding for future work.

\section{Acknowledgements}

We would like to thank D.~Berwick-Evans, C.~Closset, I.~Huq-Kuruvilla, S.~Katz, Y.~P.~Lee, I.~Melnikov, and X.~Yu for useful discussions.
E.S.~was partially supported by NSF grant PHY-2310588.

\end{document}